# Can the combination of in situ differential impedance spectroscopy and $^{27}$Al NMR detect incongruent zeolite crystallization?

*Dries Vandenabeele[a]‡, Nikolaus Doppelhammer[a]‡, Sambhu Radhakrishnan[a,b], Vinod Chandran C.[a,b], Berhard Jacoby[c], Christine Kirschhock[a], Eric Breynaert[a,b]\**

*a* Centre for Surface Chemistry and Catalysis – Characterization and Application Team (COK-KAT), KU Leuven, Celestijnenlaan 200F Box 2461, 3001-Heverlee, Belgium

*b* NMRCoRe - NMR/X-Ray platform for Convergence Research, KU Leuven, Celestijnenlaan 200F Box 2461, 3001-Heverlee, Belgium

*c* Institute for Microelectronics and Microsensors - Johannes Kepler University Linz, Altenberger Straße 69, 4040 Linz

KEYWORDS: Zeolite Crystallization – DIS – in situ NMR – HSIL

ABSTRACT

Crystallizing zeolites with isotropic properties is critical to the chemical industry but can be extremely challenging as small deviations in the synthesis conditions can have extreme effects on the final products. Easily implemented in-situ monitoring systems could make a real difference, but very few experimental methodologies cater to the specific needs of applications relying on harsh, hyper-alkaline conditions involving multiphasic systems such as Hydrated Silicate Ionic Liquids. Differential impedance spectroscopy (DIS) promises to enable such studies. It remains highly accurate despite possible electrode degradation or scaling. This study showcases how in-situ differential impedance measurements not only enable reliable detection of crystallization of even minimal amounts of zeolite product but also illustrates how the unique combination of in situ DIS and in situ, $^{27}$Al NMR provides insight into complex, incongruent zeolite crystallization processes.

INTRODUCTION

Mineral precipitation and dissolution-reprecipitation phenomena in alkaline silicate systems crucially impact energy production, nuclear waste management, and materials science. Silicate scaling in geothermal systems reduces efficiency, while glass dissolution and silicate precipitation affects the stability of nuclear waste repositories by impacting the stability of primary and secondary barriers. In material sciences, controlled zeolite crystallization is essential to produce materials with specific and constant (catalytic) properties for the chemical industry. Assessing mineral dissolution and precipitation in synthesis conditions requires detailed, time-resolved in situ physicochemical characterization, but very few experimental approaches are compatible with the hyper-alkaline conditions and multi-phasic systems that are typically encountered in zeolite synthesis media.

Zeolite formation has largely been investigated ex situ, characterizing solid products through the collection of samples at discrete time intervals, followed by X-ray diffraction (XRD) analysis to reveal product type, amount and degree of crystallinity[1,2]. Ex situ experiments, in general, suffer from low time resolution and artefacts related to process interruption and sample treatment.

For in situ characterization, far less common and not straightforward, nuclear magnetic resonance spectroscopy (NMR) stands out as the most versatile option for studying crystallization kinetics. Static NMR easily delivers quantitative, speciation-sensitive data on mobile and dissolved species. It allows us to study the most relevant elements (H, Si, Al, Na, Cs, C etc)[1,3–6] and provides insight into the connectivity of Si and Al species at any stage of the crystallization. Using MAS NMR, also the evolution of the solid phases can be studied. Drawbacks include low resolution in $^{29}$Si-NMR due to the low natural abundance of $^{29}$Si and long relaxation times, a high cost and access to dedicated equipment and sample holders to enable measurements in hydrothermal conditions[7]. In situ (synchrotron) X-ray scattering (HEXS/PDF, SAXS) can detect small length scales but this method is costly, requires specific expertise as well as access to synchrotron beamlines. In situ X-ray diffraction is insensitive to the early stages of zeolite formation due to the absence of long-range order[7,8]. Other techniques like dynamic light scattering, and vibrational spectroscopy provide partial insights but have limitations related to sensitivity, particle size, or can only provide qualitative information[8–15].

The introduction of moving electrode electrochemical impedance spectroscopy (MEEIS) offers a promising avenue for in situ analysis of zeolite crystallization and mineral precipitation/dissolution in general[16–18]. Zeolite formation, driven by the condensation of partly deprotonated aluminosilicates, generates hydroxide ions[19,20], suggesting pH monitoring to track crystallization. Unlike pH, electrical conductivity remains a sensitive metric in extremely alkaline or acidic conditions, rendering it a more reliable indicator in harsh chemical conditions. Despite its sensitivity, accurately measuring conductivity in a highly alkaline and reactive environment is challenging. Conventional sensors employing static electrodes face challenges like fouling and passivation, limiting their application for long-term in situ studies. "Contactless" toroidal conductivity sensors mitigate these issues but require large sample volumes and consequently are impractical for laboratory applications[21]. With MEEIS, impedance spectra are recorded at different inter-electrode distances. The conductivity can then easily be derived from the impedance change as function of the electrode distance. This method overcomes conventional issues, enabling highly accurate, long-term in situ measurements in corrosive media, like zeolite-forming liquids[16,18].

The value of conductivity measurements in zeolite research crucially depends on the synthesis route. Traditional hydrothermal methods involve alkaline aluminosilicate gels, which are heterogeneous, viscous, and opaque, complicating data interpretation due to sample aging, undissolved phases, and gradients. Sample heterogeneity complicates the isolation of local processes directly related to crystallization as phenomena such as reorganization of a gel phase by dissolution and reprecipitation reactions can affect the ionic conductivity of the system without immediately contributing to the crystallization process[21,22]. In extremely viscous systems, data acquisition becomes especially challenging as the accurate movement of the moving electrode in the MEEIS implementation of DIS might become impaired. Hydrated Silicate Ionic Liquids (HSILs) has been identified as an optimal alternative[23]. Composed of alkali silicate hydrates, HSILs feature minimal water content and highly deprotonated silica oligomers interacting closely with alkali cations[23–25]. These liquids are void of undissolved secondary phases, ensuring a single homogenous and transparent phase. Upon Al addition, the liquid turns from a stable state to a liquid supersaturated in aluminosilicate oligomers, yielding various zeolites, some even at room temperature[26,27]. Keeping the aluminate concentration and corresponding zeolite yield low, HSIL-based synthesis systems can be designed as homogeneous liquids with comparatively low complexity. The low yield implies the composition and properties of the directly observable liquid phase (pH, viscosity, conductivity…) are only marginally affected by crystallization. This, however, also means that only subtle changes in conductivity can be expected, requiring very high resolution and accuracy as delivered by MEEIS[18].

This research explores crystallization profiles obtained through MEEIS conductivity measurements, choosing pollucite (Cs-ANA) for its moderate crystallization temperatures and manageable synthesis times at sub-100°C conditions. By comparing the profile with synthesis curves obtained by in situ static NMR, we were able to investigate its incongruent formation behavior. Small but distinct differences in the crystallization profiles obtained by both methods were traced back to the different species actively measured by both techniques, highlighting the complementary nature of their respective data.

EXPERIMENTAL SECTION

1. *Preparation of HSIL-based synthesis liquids*

Cs-HSIL was synthesized as previously described[24], hydrolysing tetraethyl orthosilicate (TEOS) in an agitated mixture with relative molar composition: 1 TEOS (98%, Acros organics; 47.2 g) : 1 CsOH (99.5% CsOH.$H_2O$, Sigma-Aldrich; 38.1 g) : 8.7 $H_2O$ (31.5 g). During phase separation, all Si(OH)$_4$ and CsOH ends up in the dense phase[24], yielding an HSIL with relative molar composition of 1 Si(OH)$_4$ : 1 CsOH : 2.2 $H_2O$, as confirmed by gravimetric analysis[24]. Thereafter, Cs-HSIL was adjusted with caesium aluminate solution to form a final synthesis liquid with a composition of 0.5 Si(OH)$_4$ : 0.03 Al(OH)$_3$ : 1 CsOH : 9 $H_2O$. This liquid was equilibrated for 1 hour at room temperature while stirring. Subsequently, aliquots were subjected to *in situ* and *ex situ* crystallization studies. For *ex situ* characterization, 17 25g aliquots were transferred into sealed Oak-Ridge polypropylene copolymer centrifuge tubes (Nalgene) and aged in static conditions in a convection oven at 70°C. At time intervals 0 h, 1 h, 2 h, 4.2 h, 4.4 h, 4.9 h, 5.5 h, 6 h, 6.1 h, 6.9 h, 7.2 h, 8.1 h, 15.1 h, 23.2 h, 47.7 h, 71.3 h and 118.8 h (Table SI 1) one sample was removed from the oven, and immediately centrifuged at 4x10⁴ g for 15 minutes. The obtained pellet was re-dispersed in ultra-pure water and resubjected to centrifugation, exchanging the supernatant with ultra-pure water until the pH of the supernatant was below 9. Hereafter, the product was dried at 60°C for 3 days. The supernatants obtained after the initial centrifugation (mother liquors) were stored in at 4°C to inhibit further reaction.

## 2. In-situ conductivity measurements

*In situ c*onductivity experiments were performed with a custom setup employing the method of Moving Electrode Electrochemical Impedance Spectroscopy (MEEIS)[17]. This concept was developed for accurate *in situ* conductivity measurements in corrosive environments, such as the crystallization of zeolites in HSIL-based synthesis media[16,18]. Impedance spectra were recorded at 11 equally spaced inter-electrode distances between 4 and 8 cm, each at 30 logarithmically-spaced frequency points between $10^3$ and $10^6$ Hz. MEEIS scans were repeated every 15 minutes, with a scan time of approximately 120 seconds. Given the crystallization times of the chosen systems (> 2h), conductivity during data acquisition can be assumed constant. The sampling rate was adjusted for additional experiments at 40, 60, 80 and 90°C (Table SI 2).

Despite the presence of a silicone sealing ring where the shaft of the moving electrode enters the measurement volume, the possibility of sample evaporation could not be ignored. It was verified using non-reacting HSIL liquids that sample evaporation manifests in a minor, linear deviation in conductivity (Figure SI 4) To account for any, ever so slight sample evaporation, the conductivity data recorded during the crystallization experiments were therefore corrected by subtraction of a linear function, $y = ax + b$, with coefficients $a$ and $b$ determined from the conductivity data recorded well after the crystallization had ended. At this stage, all changes the Ion Activity Product (IAP) and thus also in the conductivity can almost exclusively be attributed to the impact of water evaporation on the relative fractions of ion paired versus ionic species.

3.  **In situ $^{27}$Al-NMR**

$^{27}$Al and $^{29}$Si NMR spectra were acquired on a Bruker 500 MHz Avance III spectrometer at a Larmor frequency of 500.87 MHz for $^1$H, 130.51 MHz for $^{27}$Al, and 99.5 MHz for $^{29}$Si, using with a 10 mm Si and Al – free BBO probe head (Bruker Biospin). For *in situ* $^{27}$Al measurements at 70 °C, the samples were filled in a sample cell designed for high pressure, high temperature NMR spectroscopy[28]. The sample cell consist of a 5 mm sapphire tube ($Al_2O_3$ single crystal; SP Wilmad-LabGlass; WG-507-7 series) with outer diameter (OD) 4.92 ± 0.05 mm, inner diameter (ID) 3.4 ± 0.1 mm and a total length of 178 mm connected to a section of a PEEK high-pressure liquid chromatography (HPLC). The head was sealed off to conserve mass-balance by avoiding water evaporation, and the sample was heated to the 70 °C. Spectra were acquired with a Hahn-echo pulse sequence, implementing an 22 kHz RF pulse. Time evolution of the $^{27}$Al speciation and quantification was performed until steady-state was observed (10h in this case). Each spectrum in the time series consisted of 288 transients with a relaxation delay of 1s, accounting for 5 min duration for each spectrum. The spectra were referenced to primary reference, 0.1 M $Al(NO_3)_3$ in $D_2O$. Spectral integration was performed with Bruker Topspin 4.0.9 software and spectral deconvolution was performed with DMFIT software[29].$^{29}$Si NMR spectra of the HSIL samples were also acquired at 22 °C in 3 conditions: (i) as-made HSIL before aluminate addition labelled as 'No Aluminium'; (ii) HSIL + aluminate labelled 'Before Crystallization', and (iii) HSIL synthesis media after centrifugation of the solid zeolite product, labelled 'After Crystallization'. The

measurements were performed with a Si-free 10 mm PTFE-FEP NMR tube liner (Wilmad-LabGlass). The spectra were acquired with 320 transients with a relaxation delay of 60 s, a 90° radio-frequency pulse of 22 kHz at 22 °C. $^1$H decoupling with Waltz64 sequence with an RF pulse of 4 kHz was applied during acquisition. The spectra were referenced to 0.1 M solution of Sodium trimethylsilylpropanesulfonate (DSS) in $D_2O$.

*4. Ex situ analyses*

Room temperature powder diffraction patterns were recorded on a STOE STADI MP diffractometer with Cu $K_{\alpha 1}$ radiation, focusing Ge(111) monochromator and linear position sensitive detector in Debye–Scherrer geometry. Selected samples of the time series were measured at BM01 end station of Swiss-Norwegian Beamlines at ESRF, Grenoble, France[30]. The wavelength was set to 0.7171 Å and the data was recorded with a Pilatus2M area detector. Azimuthal integration was performed by the beamline BUBBLE software[30], and additional analysis is performed with Profex software[31].

Characterization of the synthesized zeolites was performed by $^{27}$Al MAS NMR spectroscopy in Bruker 500 MHz Avance III spectrometer equipped with a 4 mm H/X/Y triple resonance magic angle spinning (MAS) probe head. The samples were packed in a 4 mm $ZrO_2$ rotor and spun at 15 kHz. The spectra were acquired with 2048 transients with a relaxation delay of 1s, RF pulse of 120 kHz, 56 kHz $^1$H decoupling with spinal64 sequence during acquisition. The spectra were referenced to $^{27}$Al resonance of 0.1 M $Al(NO_3)_3$ in $D_2O$ at 0 ppm. High-resolution scanning electron microscopy (SEM) images were recorded on a Nova NanoSEM (FEI, Hillsboro, OR). Chemical analysis of the solids (Si and Al content) was performed following digestion in $LiBO_2$ at 1000°C and dilution in HNO₃, on a Varian 720-Es ICP-OES instrument with cooled cone interphase and oxygen-free optics. The mother liquids were diluted 500x before they were measured on the same Varian instrument.

RESULTS AND DISCUSSION

In situ conductivity measurements of an HSIL-based zeolite crystallization, with composition 0.5 Si(OH)$_4$ : 0.03 Al(OH)$_3$ : 1 CsOH : 9 H$_2$O at 70°C, yields a sigmoidal profile, showing a 7% increase in conductivity (from 0.451 to 0.483 S/cm) throughout crystallization (Figure 1). In sol-gel zeolite synthesis, conductivity initially drops as silica-rich units dissolve and deprotonate, consuming OH⁻ ions and forming less mobile (alumino-)silicate species, decreasing conductivity[21,32]. However, HSILs, being homogeneous liquids at equilibrium before heating, do not exhibit such "aging" effects, avoiding the inconsistencies seen in gel systems due to false environments or initial dissolution in heterogeneous media[33]. Equilibrating the HSIL synthesis system at 70°C, the conductivity sharply rises after about 90 minutes, signaling zeolite formation[16]. As the system crystallizes, T-O-T bonds are formed by oxolation, which involves the release of mobile hydroxide. To maintain electro-neutrality, for each OH⁻ also a Cs⁺ ion is released[19,20]. The resulting conductivity increase therefore corresponds to the crystallization behaviour and is thus linked to the concentration of growth units and supersaturation.

Sigmodal crystallization curves such as shown in Figure 1 include an induction, a growth and a plateau phase. They can be described analytically using the Finke–Watzky two-step kinetic model incorporating slow, continuous nucleation and fast autocatalytic growth[34]. In the mathematical model, maximum acceleration signifies the transition from the induction period (3.27 h) to the exponential growth phase, with the inflection point on the sigmoid curve ($t_{0.5}$=3.87 h) marking the peak growth rate (0.018 S/h.cm). after which the growth rate slows down by increasing depletion of growth units. After approximately 8 h, the conductivity curve in Figure 1 reached a plateau, indicating the end of crystal growth and the start of a dynamic equilibrium between the crystalline and liquid phases. Growth slows as growth units deplete, and by around 8 hours, the conductivity in Figure 1 plateaus, indicating crystal growth cessation and the onset of dynamic equilibrium between crystalline and liquid phases. This equilibrium, marked by ongoing dissolution and precipitation, maintains the aluminosilicate content in the crystalline phase steady, with net no new T-O-T bonds forming, thus stabilizing liquid phase conductivity. These results suggests that liquid phase conductivity in contact with the growing crystalline phase can serve as proxy for the crystallization process.

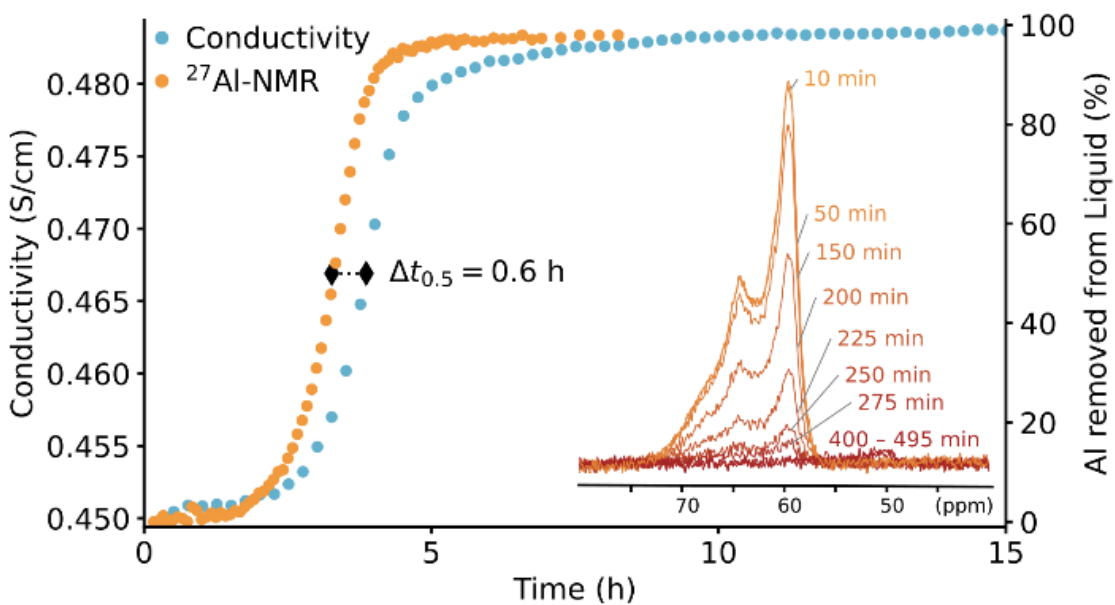

Figure 1: Time evolution of the MEEIS conductivity and fractional removal of Al from the synthesis solution. The inset shows the evolution of the quantitative $^{27}$Al NMR spectra as a function of time, demonstrating a minimal change in speciation over time.

To validate this hypothesis, the same crystallization was monitored with static in situ $^{27}$Al NMR, quantitatively measuring the removal of Al from the liquid phase. Previous ex situ studies have demonstrated that virtually all Al removed from the HSIL liquid is incorporated into the zeolitic phase[33], a finding this study also supports (Figure SI 3). Consequently, the fractional concentration of Al removed from the liquid {Al}=1-([Al]/[Alt0]) is a quantitative measure for crystal growth (Figure 1). The curve in Figure 1 was derived by integrating all quantitative $^{27}$Al spectra over time, calculating the remaining fractional Al concentration after normalizing to the initial spectrum. This analysis revealed a 98% transfer of Al into the solid phase.

Deconvolution and peak assignment of the $^{27}$Al NMR (Figure SI 8), revealed the exclusive presence of $q^4$, $q^3$ and $q^2$ Al species (46%, 26.6% and 27.4% respectively). Consistent with expectations for systems rich in silicate and highly alkaline[35], the results indicate that monomeric aluminate ion concentrations always remained below the limit of detection. Throughout the crystallization process, the relative speciation remained constant owing to the low yield, causing only minimal changes in the crystallization medium. Analogously, the silicon speciation, assessed with $^{29}$Si-NMR, experienced minimal changes (Figure SI 10).

Direct comparison between the synthesis curves from MEEIS and $^{27}$Al NMR spectroscopy reveals a 36-minute shift (9% of the total crystallization time) in the inflection points of the sigmoid curves, with NMR at 3.27 h and MEEIS at 3.87 h. COMSOL simulations indicate this effect is partly due to varying transient temperature effects during the initial heating stage in both setups (Figure SI 7). Additional experiments in which the crystallization temperature was varied show a highly similar synthesis profile, yet small variations in temperature could lead to drastic changes in the $t_{0.5}$, as is highlighted in SI.

After re-normalization intensity of the conductivity curve and aligning the NMR and conductivity crystallization profiles at the time of their inflection points, both profiles exhibit a remarkable degree of agreement up to 80% of the reaction progress (Figure 2). This strongly supports the hypothesis that the conductivity changes of the liquid phase correlate with the formation of the solid product over time in these systems. The difference in crystallization profiles towards the end of the synthesis is explained by differences in the measurement techniques, with each capturing different aspects of the phase transition process. Conductivity reflects all ionic interactions and is mainly affected by hydroxide release during (alumino-)silicate bond formation. Conversely, $^{27}$Al NMR probes the concentration of liquid-borne aluminium, providing no information about silicate incorporation in the zeolite framework. Reconstruction of the crystallization profiles from liquid state $^{27}$Al NMR data relies on the assumption that zeolite formation proceeds at a constant solid silicon-to-aluminium (Si/Al) ratio, a rough approximation. Ex situ elemental analysis shows the Si/Al ratio in the solid phase increases from about 2.1 to 2.6 throughout crystallization (Figure 3 (a)), indicating a higher silicon incorporation per aluminium into the zeolite as the reaction progresses. Additional $^{27}$Al NMR analysis of the solids after 4.4h and 118h assures a negligible influence by extra-framework Al (Figure SI 3). Interestingly, the lowest Si/Al ratio corresponds most closely to the highest fractional occupancy of cation positions (Si/Al = 2.0)[36], the composition with the highest thermodynamic stability.

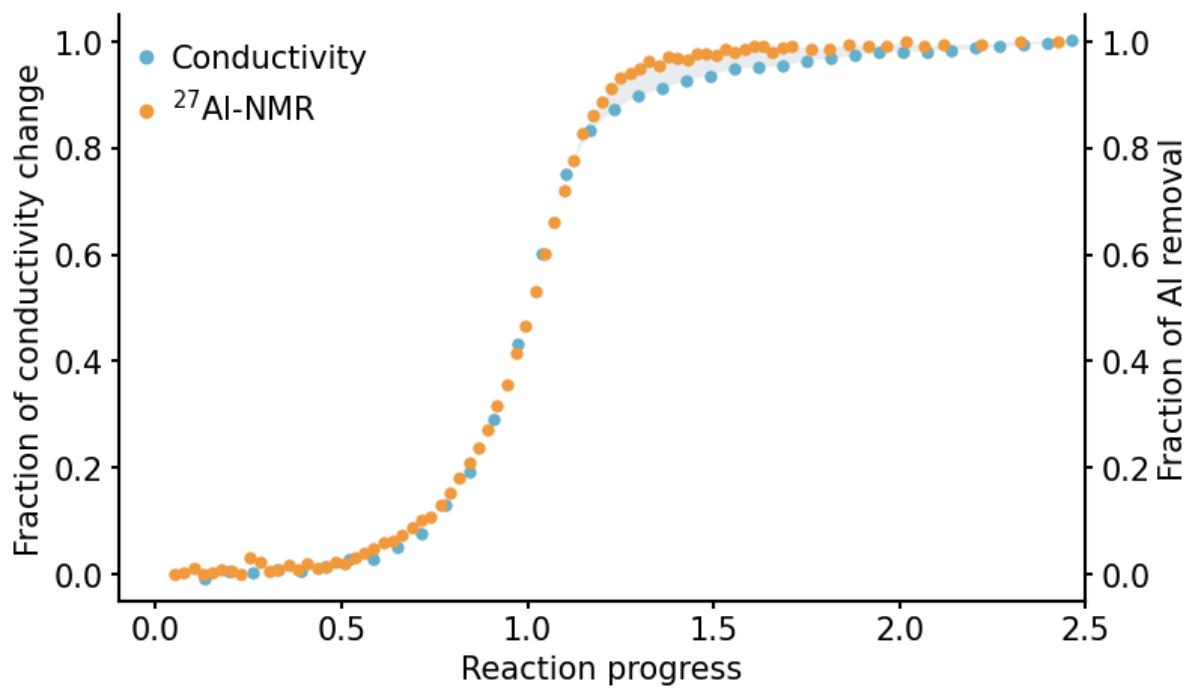

Figure 2: Comparison of normalized conductivity and Al-NMR profiles.

Yield measurements are generally not suitable for kinetic studies due to low temporal resolution and artefacts stemming from process interruptions and sample treatment[2], as highlighted in the Supplementary Information for this study. However, they are useful for examining trends in the properties of solids. Synchrotron powder XRD of selected samples showed a phase pure Cs-ANA product[37], with peak widths narrowing over time, indicating improved crystallinity or larger crystal sizes (Figure 3 (c,f)). Analysis with Profex software attributed the peak narrowing partly to an increase in crystallite size (Figure 3 (d)). Additionally, it was found that a higher Si/Al ratio leads to unit cell compression (Figure 3 (e)), with more siliceous compositions exhibiting smaller unit cells due to the effects of shorter binding angles and longer Al-O bonds. Electron microscopy revealed the solids as spherical particles uniformly sized between 40 to 60 nm (Figure SI 1).

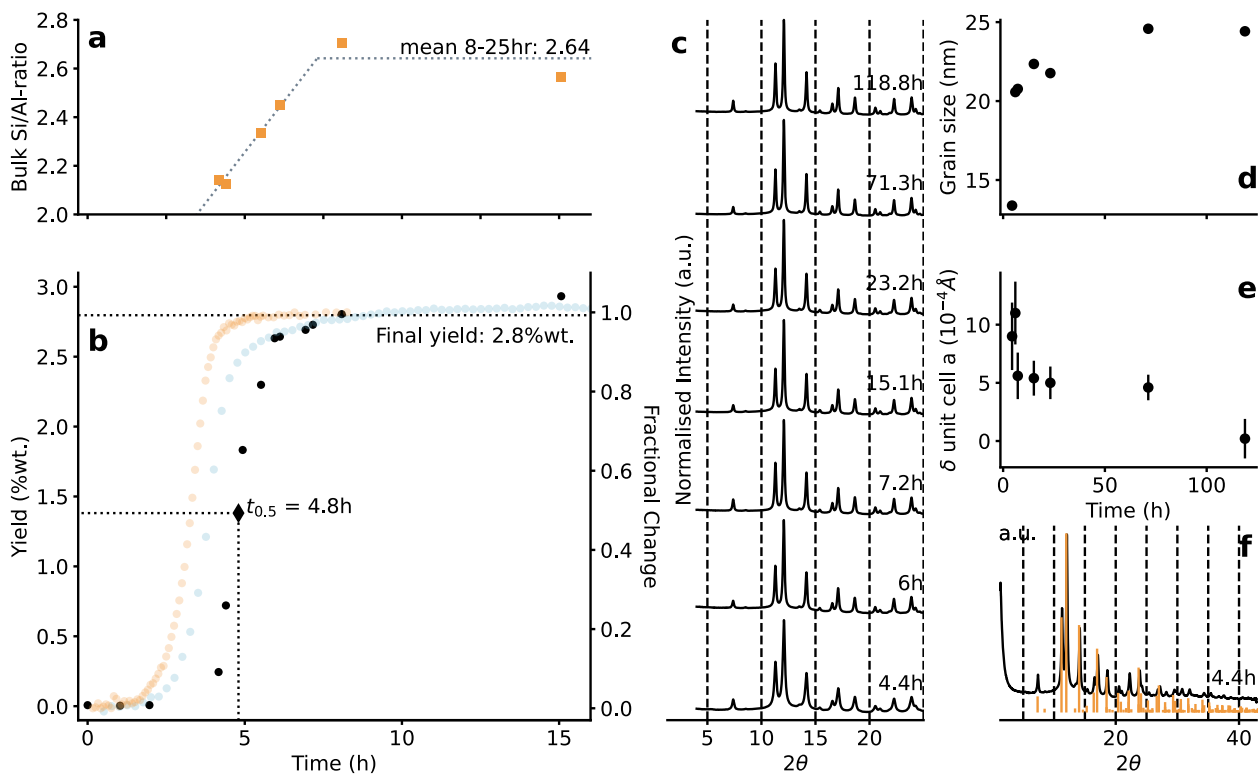

Figure 3: (a) Si/Al ratio of the solid product over time, the dotted lines are added to guide the reader. (b) Zeolite yield profile as function of time, exhibiting a sigmoidal shape. The maximal yield is 2.8 wt%, obtained after approximately 10h. The inflection point is reached at t = 4.83h. NMR and Conductivity data are added in respectively orange and blue; (c) The XRD patterns recovered as function of time, collected at wavelength 0.7171 Å; (d) The grain size of the crystallites increases over time, while the unit cell is increasingly more compressed reaching 1.3646 Å after 119 h (e). (f) Comparison of the first pattern with the reference of the RRUFF database[37], identifying the sample as phase pure pollucite. The patterns are normalised to the reflection at 11.98°.

CONCLUSION

This study showcases and validates *in situ* differential impedance measurements for detecting zeolite crystallization of. It highly sensitively detects crystallization of zeolites even from highly alkaline, highly concentrated silicate liquids. By implementing conductivity measurements in differential mode, the measurements remain highly accurate despite electrode corrosion or scaling. The DIM results were evaluated against in situ $^{27}$Al NMR, detecting Al removal from solution and subsequent conversion into crystalline Pollucite zeolite as demonstrated by synchrotron X-ray diffraction. Comparison of the synthesis profiles obtained from NMR and DIM reveal small differences, which may originate from an incongruent crystallization mechanism, resulting in an increasing Si/Al-ratio as the crystallization proceeds. These findings are supported by thorough ex situ characterization including synchrotron XRD and MAS NMR. Overall, it was demonstrated how combining these methods offers profound insights into mineral dissolution and re-precipitation mechanisms.

The presented approach holds promise to enhance the comprehension of dissolution and repricipitation mechanisms beyond zeolitic systems, where phase transformation is accompanied by the transformation of charged species.

AUTHOR INFORMATION

**Corresponding Author**

* Eric Breynaert; eric.breynaert@kuleuven.be

**Author Contributions**

‡Shared first authors with equal contributions to this work.

ACKNOWLEDGMENT

E.B., and C.E.A.K. acknowledge joined funding by the Flemish Science Foundation (FWO; G083318N) and the Austrian Science Fund (FWF) (funder ID 10.13039/501100002428, project ZeoDirect I 3680-N34). This work has received funding from the European Research Council (ERC) under grant agreement no. 834134 (WATUSO). E.B. acknowledges FWO for a "Krediet aan navorsers" 1.5.061.18N. NMRCoRe is supported by the Hercules Foundation (AKUL/13/21), by the Flemish Government as an international research infrastructure (I001321N), and by Department EWI via the Hermes Fund (AH.2016.134). The authors acknowledge the SNBL beamline (ESRF, Grenoble) for experimental support.